\documentclass[aps,superscriptaddress,amsmath,amssymb,nofootinbib,11pt]{revtex4-1}

\usepackage{titlesec}
\usepackage[titletoc,toc,page,header]{appendix}
\usepackage{tabularx}
\usepackage{graphicx}
\usepackage{amsmath}
\usepackage{geometry}
\usepackage{amssymb}
\usepackage{mathrsfs}
\usepackage{float}
\usepackage{makecell}
\usepackage{multirow}
\usepackage{latexsym}
\usepackage{color}
\usepackage{subfigure}
\usepackage{slashed}
\usepackage{hyperref}
\usepackage{lipsum}
\usepackage{threeparttable}

\def\be{\begin{eqnarray}}\def\ee{\end{eqnarray}}

\begin{document}
\title{Nuclear matter properties from chiral-scale effective theory including a dilatonic scalar meson}

\author{Lu-Qi Zhang}
\affiliation{School of Physics, Nanjing University, Nanjing, 210093, China}
\affiliation{School of Frontier Sciences, Nanjing University, Suzhou, 215163, China}
\affiliation{School of Fundamental Physics and Mathematical Sciences, Hangzhou Institute for Advanced Study, UCAS, Hangzhou, 310024, China}

\author{Yao Ma}
\email{mayao@nju.edu.cn}
\affiliation{School of Frontier Sciences, Nanjing University, Suzhou, 215163, China}
\affiliation{School of Fundamental Physics and Mathematical Sciences, Hangzhou Institute for Advanced Study, UCAS, Hangzhou, 310024, China}

\author{Yong-Liang Ma}
\email{ylma@nju.edu.cn}
\affiliation{School of Frontier Sciences, Nanjing University, Suzhou, 215163, China}
\affiliation{International Center for Theoretical Physics Asia-Pacific (ICTP-AP) , UCAS, Beijing, 100190, China}

\begin{abstract}

Chiral effective theory has become a powerful tool for studying the low-energy properties of QCD.
In this work, we apply an extended chiral effective theory---chiral-scale effective theory---including a dilatonic scalar meson to study nuclear matter and find that the properties around saturation density can be well reproduced.
Compared to the traditionally used Walecka-type models in nuclear matter studies, our approach improves the behavior of symmetry energy and the incompressibility coefficient in describing empirical data without introducing additional freedoms.
Moreover, the predicted neutron star structures fall within the constraints of GW170817, PSR J0740+6620, and PSR J0030+0451, while the maximum neutron star mass can reach about \(~3M_{\odot}\)  with a pure hadronic phase.
Additionally, we find that symmetry patterns of the effective theory significantly impact neutron star structures.
We believe that introducing this type of theory into nuclear matter studies can lead to a deeper understanding of QCD, nuclear matter, and compact astrophysical objects.

\end{abstract}

\maketitle

\allowdisplaybreaks{}

\section{Introduction}

\label{sec:Intro}

Study on nuclear matter (NM) has long been crucial for understanding both nuclear force and neutron star (NS) structures (see, e.g., Refs.~\cite{Serot:1984ey, Brown:2001nh,Fukushima:2010bq,Lattimer:2015nhk, Baym:2017whm,Ma:2019ery,Brandes:2023bob} and references therein).
The properties of NM are highly sensitive to details of nucleon interactions.
A popular and widely used approach to describe these interactions is the one-boson-exchange (OBE) model and its variants, such as the Walecka-type models~\cite{Holinde:1975vg, Erkelenz:1974uj, Nagels:1977ze, Machleidt:1989tm}.
These models typically involve \(\pi\), \(\sigma\), \(\omega\) and \(\rho\) meson exchanges, covering the effective range of nucleon forces from \(0.5\ \mathrm{fm}\) to \(2\ \mathrm{fm}\).
Utilizing these models, NM properties can be calculated using the relativistic mean field (RMF) approach~\cite{Walecka:1974qa, Wiringa:1988tp,Gao:2022klm,Gao:2024jlp}, which is the most practical and economical framework to introduce density effects.

It is recognized that Walecka-type models lack the consideration of QCD symmetry patterns, the valid region of effective operators, and theoretical errors. 
Chiral effective field theory (\(\rm{\chi EFT}\)), thanks to the pioneering works by Weinberg \cite{Weinberg:1978kz, Weinberg:1991um}, offers a powerful framework for studying nuclear forces at long ranges, anchored on QCD symmetry.
To develop a realistic model of nuclear forces, it is accepted that vector mesons---\(\rho\) and \(\omega\)--- and isoscalar-scalar meson \(\sigma\) are indispensable.
The former can be regarded as gauge fields of hidden local symmetry (HLS) in \(\rm{\chi EFT}\) ~\cite{Bando:1984ej,Bando:1987br, Harada:2003jx}.
However, the inclusion of the \(\sigma\) meson as an independent degree of freedom in \(\rm{\chi EFT}\) is not straightforward, since the fourth scalar component of the chiral four-vector is integrated out when transitioning from the linear sigma model to the nonlinear sigma model---the leading term of \(\rm{\chi EFT}\).
Considering that the lowest-lying scalar meson \(\sigma\) has a mass roughly equivalent to that of the kaon, which is a Nambu-Goldstone (NG) boson of chiral symmetry breaking in three-flavor \(\rm{\chi EFT}\) (\(\chi\rm EFT_3\)), by supposing that QCD has a nonperturbative infrared fixed point, Crewther and Tunstall \cite{Crewther:2013vea, Cata:2018wzl} proposed that the lowest-lying scalar meson can be considered as the NG boson (dilaton) of the spontaneous breaking of scale symmetry.
Consequently, \(\rm{\chi EFT}\) is extended to include the scalar meson, resulting in the scale-chiral effective theory (\(\chi \rm EFT_{\sigma}\)).

Based on the terminologies of \(\chi \rm EFT_{\sigma}\) and HLS, the \(\chi \rm EFT\) with baryons is constructed in Refs.~\cite{Lee:2015qsa,Li:2016uzn} at the leading chiral order to discuss nucleon interactions.
It is denoted as bsHLS, with `b' for baryon and `s' for scale.
In bsHLS, the potential from meson exchanges in OBE models can be reproduced by expanding bsHLS to the first order of linear field couplings but with additional symmetry considerations. By using the low-momentum potential $V_{\rm low k}$ renormalization group approach~\cite{Bogner:2003wn,Holt:2006ii}, taking the ``leading order scale symmetry (LOSS)” approximation where the trace anomaly effect enters only through the dilaton potential, which breaks scale symmetry explicitly and spontaneously, we found that the chiral-scale EFT with few parameters can successfully describe not only NM at the saturation density but also the
compact-star matter at $n=(5-7)n_0$~\cite{Ma:2019ery}.
A novel phenomena, which was not realized before, is that in nuclear matter at densities $n=(2-4)n_0$ the sound velocity of NM saturates the conformal limit $v_s^2 =1/3$, but the trace of the energy-momentum tensor does not vanish, that is, the NM exhibits a pseudo-conformal structure~\cite{Ma:2018xjw,Ma:2018qkg,Ma:2018jze}.
And the corrections to LOSS in the baryonic part have been found crucial for understanding the quenched $g_A$ value in the super-allowed Gamow-Teller transitions of heavy nuclei~\cite{Lubos:2019nik,Ma:2020tsj}.

In this work, we study NM properties using bsHLS with the RMF method. After identifying the NM properties around \(n_0\), it was found that bsHLS using the RMF method can yield reasonable results.
The data can be reproduced with the choice of \(\beta'=\frac{\partial\beta}{\partial\alpha}\sim 1\), consistent with the results of Refs.~\cite{Ma:2016nki, Shao:2022njr}, where a dilaton limit fixed point is assumed in the medium.
The combination of $f_{\chi}m_{\sigma}$ is constrained to be $\simeq 2.3 \times 10^{5}$~MeV$^2$ which is consistent with what estimated before in the skyrmion crystal approach such that physically interesting results can be obtained~\cite{Lee:2003eg,Park:2008zg}.
By comparing the symmetry energy \(E_{\rm sym}\) and incompressibility coefficient \(K(n)\) obtained from our bsHLS and Walecka-type models, we find that the bsHLS can make the symmetry energy stiff at subsaturation density but soft at intermediate densities to meet the constraints of GW170817~\cite{LIGOScientific:2018cki} and the neutron skin thickness of $^{208} \rm Pb$~\cite{Reed:2021nqk} simultaneously, without introducing additional freedoms such as the \(\delta\) meson in Refs.~\cite{Li:2022okx,Ma:2023eoz}.
In addition, we find that the incompressibility coefficient of bsHLS surges at intermediate densities, while the Walecka-type models exhibit gentler behaviors, resulting in a better description of NS structures using bsHLS.
The maximum mass of NS can nearly reach \(3M_{\odot}\) with a pure hadron phase, meeting the constraints of GW170817~\cite{LIGOScientific:2018cki, LIGOScientific:2018hze} and PSR J0740+6620~\cite{Miller:2021qha,Riley:2021pdl}, whereas Walecka-type models can only reach \(2M_{\odot}\) within these constraints, as analyzed in Ref.~\cite{Guo:2023mhf}.
We find that this is due to the kink behavior of \(\sigma\) expectations at intermediate densities, induced by the nonlinear realization of scale symmetry.
Moreover, the value of \(\beta'\), behavior of \(\langle\chi\rangle^*\) and Brown-Rho scaling (B-R scaling) at different densities can significantly affect NS structures, indicating a relationship between QCD symmetry patterns and macroscopic phenomena.

This article is organized as follows: In Sec.~\ref{sec:bsHLS}, we introduce the theoretical framework of bsHLS and the equation of state (EoS) of NM under the RMF approximation. In Sec.~\ref{sec:Phen}, we provide a phenomenological analysis by pinning the experimental data of nuclear matter and neutron star observations. A comparison with the Walecka-type models is also made. Our summary and discussion are presented in Sec.~\ref{sec:Conc}.

\section{bsHLS in nuclear medium}

\label{sec:bsHLS}

In this section, we establish the theoretical framework of bsHLS and derive the equation of state (EoS) for NM within the relativistic mean field (RMF) approach.
Focusing on NM only composed of nucleons, we restrict the bsHLS Lagrangian to the two-flavor case, involving only \(u\) and \(d\) quarks.
The leading order bsHLS Lagrangian \(\mathcal{L}\) can be decomposed into the baryonic part \(\mathcal{L}_B\) and the mesonic part \(\mathcal{L}_{M}\):
\begin{equation}
	\label{eq:L}
    \mathcal{L}=\mathcal{L}_M+\mathcal{L}_B\ .
\end{equation}

The mesonic part \(\mathcal{L}_M\), consisting of \(\sigma\), \(\rho\), \(\omega\) and \(\pi\) mesons, is constructed as follows~\cite{Li:2016uzn},
\be
\label{eq:LM}
\mathcal{L}_M & = & \ f^2_\pi\Phi^2\left[h_1+\left(1-h_1\right)\Phi^{\beta'}\right]{\rm Tr}\left(\hat{\alpha}^\mu_\perp\hat{\alpha}_{\mu\perp}\right) \nonumber\\
& &{} +\frac{m^2_\rho}{g^2_\rho}\Phi^2\left[h_2+(1-h_2)\Phi^{\beta'}\right]{\rm Tr}\left(\hat{\alpha}^\mu_\parallel\hat{\alpha}_{\mu\parallel}\right) \nonumber\\
& &{} +\frac{1}{2}\left(\frac{m^2_\omega}{g^2_\omega}-\frac{m^2_\rho}{g^2_\rho}\right)\Phi^2\left[h_3+\left(1-h_3\right)\Phi^{\beta'}\right]{\rm Tr}\left(\hat{\alpha}^\mu_\parallel\right){\rm Tr}\left(\hat{\alpha}_{\mu\parallel}\right) \nonumber\\
& &{} +\frac{1}{2}\left[h_4+\left(1-h_4\right)\Phi^{\beta'}\right]\partial_\mu\chi\partial^\mu\chi+\frac{f^2_\pi}{4}\Phi^
   {3-\gamma_m}{\rm Tr}\left(\mathcal{M}U^\dagger+U\mathcal{M}^\dagger\right)+h_5\Phi^4+h_6\Phi^{4+\beta'} \nonumber\\
& &{} -\frac{1}{2g^2_\rho}\left[h_7+\left(1-h_7\right)\Phi^{\beta'}\right]{\rm Tr}\left(V_{\mu\nu}V^{\mu\nu}\right)-\frac{1}{2g^2_0}\left[h_8+\left(1-h_8\right)\Phi^{\beta'}\right]{\rm Tr}\left(V_{\mu\nu}\right){\rm Tr}\left(V^{\mu\nu}\right) \ , \nonumber\\
\ee
where $V_{\mu \nu}=\partial_\mu V_\nu-\partial_\nu V_\mu-i\left[V_\mu, V_\nu\right]$, $V_\mu=\frac{1}{2}\left(g_\omega \omega_\mu+g_\rho \rho_\mu^a \tau^a\right)$.
$m_\rho$, $m_\omega$ and \(\mathcal{M}=m_{\pi}^2 I_{2\times2}\) are the masses of $\rho$, $\omega$ and \(\pi\) mesons, respectively.
The dilaton field $\chi$ is introduced as a nonlinear representation $\chi=f_\chi\Phi=f_\chi \exp \left(\sigma / f_\chi\right)$, and \(\beta'\) accounts for the anomalous dimension of gluon field operators, representing the deviation from the IR fixed point.
According to Ref.~\cite{Li:2017hqe}, $h_i$s, except $h_5$ and $h_6$, are chosen to be $1$ for simplification and the anomalous dimension \(\gamma_m\) is simply taken to be $1$ (denoted as LOSS).
$h_5$ and $h_6$ are constrained by the saddle point equations, 
\be
\label{eq:sp}
& & 4 h_5+\left(4+\beta^{\prime}\right) h_6+2 m_\pi^2 f_\pi^2=0\ , \nonumber\\
& & 12 h_5+\left(4+\beta^{\prime}\right)\left(3+\beta^{\prime}\right) h_6+2 m_\pi^2 f_\pi^2=-m_\sigma^2 f_\chi^2\ .
\ee
In Lagrangian~\eqref{eq:LM}, pions are introduced as a nonlinear field $\xi=\sqrt{U}=e^{i \frac{\pi}{f_{\pi}}}$, where $\pi=\pi^a \tau^a$ with $a=1,2,3$ representing the isospin indices.
Its covariant derivative is defined as
\begin{equation}
    D_\mu \xi=\left(\partial_\mu-i V_\mu\right) \xi =\left[\partial_\mu-i\frac{1}{2}\left(g_\omega\omega_\mu+g_\rho\rho_\mu^a\tau^a\right)\right] \xi\ .
\end{equation}
The Murrer-Cartan 1-forms $\hat{\alpha}_{\perp}^\mu$ and $\hat{\alpha}_{\|}^\mu$ are defined as
\begin{equation}
    \hat{\alpha}_{\perp,\| }^\mu=\frac{1}{2 i}\left(D^\mu \xi \cdot \xi^{\dagger}\mp D^\mu \xi^{\dagger} \cdot \xi\right)\ .
\end{equation}

The baryonic part Lagrangian \(\mathcal{L}_B\) is written as~\cite{Li:2016uzn}
\be
\label{eq:LB}
\mathcal{L}_B & = & \left[g_1+(1-g_1)\Phi^{\beta'}\right]\bar{N} i \gamma_\mu D^\mu N-\left[g_2+\left(1-g_2\right)\Phi^{\beta'}\right]m_N \Phi \bar{N} N \nonumber\\
& &{} +\left[g_A C_A+g_A\left(1-C_A \right)\Phi^{\beta^{\prime}}\right] \bar{N} \hat{\alpha}_{\perp}^\mu \gamma_\mu \gamma_5 N+\left[g_{V_\rho} C_{V_\rho}+g_{V_\rho}\left(1-C_{V_\rho}\right) \Phi^{\beta^{\prime}}\right] \bar{N} \hat{\alpha}_{\|}^\mu \gamma_\mu N \nonumber\\
& &{} +\frac{1}{2}\left[g_{V_0} C_{V_0}+g_{V_0}\left(1-C_{V_0}\right) \Phi^{\beta^{\prime}}\right] \operatorname{Tr}\left[\hat{\alpha}_{\|}^\mu\right] \bar{N} \gamma_\mu N\ ,
\ee
where \(N\) is the iso-doublet of baryon field, and $g_1$ and $g_2$ are set to $1$ as suggested in Ref.~\cite{Li:2017hqe}.
For convenience, we introduce the combinations of the parameters:
\be
& & g_{\omega NN}=\frac{1}{2}\left(g_{V_p} C_{V_p}+g_{V_0} C_{V_0}-1\right) g_\omega\ ,\nonumber\\
& & g_{\rho NN}=\frac{1}{2}\left(g_{V_p} C_{V_p}-1\right) g_\rho\ ,\nonumber\\
& & g^{SSB}_{\omega NN}=\frac{1}{2}\left[g_{V_\rho}\left(1-C_{V_\rho}\right)+g_{V_0}\left(1-C_{V_0}\right)\right]g_\omega\ ,\nonumber\\
& & g^{SSB}_{\rho NN}=\frac{1}{2}\left[g_{V_\rho}\left(1-C_{V_\rho}\right)\right]g_\rho\ .
\ee

By regarding the NM as homogeneous matter, the RMF approximation can be applied. Using Lagrangian~\eqref{eq:L}, the EOMs of \(\omega\) and \(\rho\) can be obtained as
\be
\label{eq:EOMv}
& & m^2_\omega\Phi^2\omega-\left[g_{\omega NN}+g^{SSB}_{\omega NN}\left(\Phi^{\beta'}-1\right)\right]\left(\rho_n+\rho_p\right)=0\ ,\nonumber\\
& & m^2_\rho\Phi^2\rho  -\left[g_{\rho NN}+g^{SSB}_{\rho NN}\left(\Phi^{\beta'}-1\right)\right]\left(\rho_p-\rho_n\right)=0\ ,
\ee
where \(\rho\) and \(\omega\) fields are denoted, respectively, as \(\rho\) and \(\omega\) for brevity.
Similarly, the EOM of \(\sigma\) field is derived as
\be
m^2_\omega\omega^2\Phi+m^2_\rho\rho^2\Phi & = & \frac{m^4_N\Phi^3}{\pi^2}\left[F\left(\frac{k_p}{m_N\Phi}\right)+F\left(\frac{k_n}{m_N\Phi}\right)\right]\nonumber\\
& &{} -2f^2_\pi m^2_\pi\Phi-4h_5\Phi^3-\left(4+\beta'\right)h_6\Phi^{3+\beta'} \nonumber\\
& &{} + g^{SSB}_{\omega NN}\beta'\Phi^{\beta'-1}\omega\left(\rho_p+
		\rho_n\right)+g^{SSB}_{\rho NN}\beta'\Phi^{\beta'-1}\rho\left(\rho_p-\rho_n\right)\ ,
\ee
where $\frac{m_N^3 \Phi^3}{\pi^2}F\left(\frac{k_{p(n)}}{m_N \Phi}\right)$ refers to scalar density $\langle\bar{p} p\rangle$ or $\langle\bar{n} n\rangle$ and \(k_{p(n)}\) is the Fermi momentum of nucleons at zero temperature.

The energy density can be obtained via Lagrangian~\eqref{eq:L} with the solutions of EOMs mentioned above
\be
 \langle\mathcal{H}\rangle & = & {-i\left\langle\bar{N} \gamma_i \partial^i N\right\rangle+m_N \Phi\langle\bar{N} N\rangle}  +\frac{1}{2} m_\omega^2 \Phi^2 \omega^2+\frac{1}{2} m_\rho^2 \Phi^2 \rho^2 -f_\pi^2 m_\pi^2 \Phi^2 -h_5 \Phi^4-h_6 \Phi^{4+\beta^{\prime}} \nonumber\\
& = & \frac{m_N^4 \Phi^4}{\pi^2}\left[f\left(\frac{k_p}{m_N \Phi}\right)+f\left(\frac{k_n}{m_N \Phi}\right)\right]+\frac{1}{2} m_\omega^2\Phi^2 \omega^2+\frac{1}{2} m_\rho^2 \Phi^2 \rho^2 \nonumber\\
& &{} -f_\pi^2 m_\pi^2 \Phi^2-h_5 \Phi^4-h_6 \Phi^{4+\beta^{\prime}}\ ,
\ee
where \(f\left(\frac{k_p(n)}{m_N \Phi}\right)=\int_0^{\frac{k_p(n)}{m_N}} x^{\prime 2} \sqrt{1+x^{\prime 2}} {\mathrm{d}} x^{\prime}\).
It should be noted that the nonlinear Lagrangian~\eqref{eq:L} will lead to nonzero constant vacuum energy \(\mathcal{E}_0=-f_\pi^{2} m_\pi^2-h_5-h_6\), which is neglected in our equation of state (EOS) calculations.

The above bsHLS~\eqref{eq:L} is constructed in a matter-free space. When applying it to a medium, it's natural to expect that the parameters in the Lagrangian should be changed by medium, here density.
We implement this density effect via Brown-Rho scaling (B-R scaling)~\cite{Brown:1991kk, Brown:2001nh}, and refer to this density effect as intrinsic density dependence (IDD).
Explicitly, the parameters in Lagrangian~\eqref{eq:L} scale as
\begin{equation}
    \begin{aligned}~\label{eq:br-scaling}
       \frac{m_{\rho(\omega,N)}^*}{m_{\rho(\omega,N)}} \approx \frac{f_{\pi}^*}{f_{\pi}} \approx\Phi^*\ ,\quad \frac{m_\sigma^*}{m_\sigma} \approx\left(\Phi^*\right)^{1+\frac{\beta^{\prime}}{2}}\ .
    \end{aligned}
\end{equation}
A possible choice of \(\Phi^*\) is \(1/(1+r \frac{\rho_n+\rho_p}{n_0})\).
Pion-nuclei bound state data~\cite{Kienle:2004hq} indicates \(r\approx 0.2\), but we set it as a free parameter to fit the NM properties in this work.
Chiral dynamics indicates that pion mass is not changed by medium, so we set \(m_{\pi}^*/m_{\pi}\approx 1\).
The saddle point equation~\eqref{eq:sp} leads to
\begin{equation}
    \begin{aligned}
    h_5^*=&\frac{-2\left(2+\beta^{\prime}\right) m^{*2}_\pi f^{*2}_\pi+m^{*2}_\sigma f^{*2}_\chi}{4 \beta^{\prime}}\ ,\quad h_6^*=& \frac{4 m^{*2}_\pi f^{*2}_\pi-m^{*2}_\sigma f^{*2}_\chi}{\left(4+\beta^{\prime}\right) \beta^{\prime}}\ .
    \end{aligned}
\end{equation}

After the above discussions, we finally obtain the energy density for phenomenological analysis as
\be
\mathcal{E} & = & \frac{m_N^{* 4} \Phi^4}{\pi^2}\left[f\left(\frac{k_p}{m_N^* \Phi}\right)+f\left(\frac{k_p}{m_N^* \Phi}\right)\right]+\frac{1}{2} m_\omega^{* 2} \Phi^2 \omega^2+\frac{1}{2} m_\rho^{* 2} \Phi^2 \rho^2 \nonumber\\
& &{} -f_\pi^{* 2} m_\pi^2 \Phi^2-h_5^* \Phi^4-h_6^* \Phi^{4+\beta^{\prime}}-\mathcal{E}_0^*\ .
\ee

\section{Phenomenological analysis}

\label{sec:Phen}

In the phenomenological analysis, we choose vacuum values \(f_{\pi}=92.4~\rm MeV\), \(m_N=939~\rm MeV\), \(m_{\pi}=140~\rm MeV\), \(m_{\omega}=783~\rm MeV\) and \(m_{\rho}=765~\rm MeV\)~\cite{ParticleDataGroup:2022pth}.
The free parameters are \(M_{\sigma}=m_{\sigma}f_{\chi}\), \(\beta'\), \(r\), \(g_{\omega NN}\), \(g_{\rho NN}\), \(g^{SSB}_{\omega NN}\) and \(g^{SSB}_{\rho NN}\), which can be estimated by the properties of NM around saturation density \(n_0\).
The calculated results of NM properties are listed in Table~\ref{tab:nuclear-matter}, and the corresponding low energy constants (LECs) are given in Table~\ref{tab:para}.
It can be seen that NM properties obtained from both bsHLS-L and bsHLS-H are consistent with empirical values.
\begin{table}[tbh]\small
    \centering
    \caption{
        The properties of nuclear matter: \(e_0\) is the binding energy of nucleon at \(n_0\), $E_{\mathrm{sym}}(n)=\left.\frac{1}{2} \frac{\partial^2 E(n, \alpha)}{\partial \alpha^2}\right|_{\alpha=0}$ is the symmetry energy, $\left.K_0=9 n^2 \frac{\partial^2 E(n, 0)}{\partial n^2}\right|_{n=n_0}$ is the incompressibility coefficient, $\left.J_0=27 n^3 \frac{\partial^{3} E(n, 0)}{\partial n^3}\right|_{n=n_0}$ is the skewness coefficient and $L(n)=3 n \frac{\partial E_{\mathrm{sym}}(n)}{\partial n}$ is the symmetry energy density slope.
        \(n_c\approx 0.11\mathrm{fm^{-3}} \) is subsaturation cross density.
        Two sets of predictions are shown: bsHLS-L refers to the case, where the surge of \(K(n)\) located at lower density regions, and bsHLS-H refers to the higher density case.
        \(n_0\) is in the unit of \(\rm fm^{-3}\), and the others are in the unit of \(\rm MeV\).
    }
 ~\label{tab:nuclear-matter}
        \begin{tabular}{@{}cccc}
            \hline
            \hline
            & Empirical & bsHLS-L & bsHLS-H \\
            \hline
            $n_0$ & $0.155\pm0.050$~\cite{Sedrakian:2022ata} & 0.159 & 0.159 \\
            \hline
            \(e_0\) & $-15.0\pm1.0$~\cite{Sedrakian:2022ata} & $-16.0$ & $-16.0$ \\
            \hline
            $K_0$ & $230\pm30$~\cite{Dutra:2012mb} & 232 & 284\\
            \hline
            $E_{\mathrm{sym}}(n_c)$ & $22.4\pm2.3$~\cite{Chen:2011ek,Liu:2010ne} & 20.8 & 20.9\\
            \hline
            $E_{\mathrm{sym}}(n_0)$ & $30.9\pm1.9$~\cite{Lattimer:2012xj} & 30.5 & 29.2\\
            \hline
            $E_{\mathrm{sym}}(2n_0)$ & $46.9\pm10.1$~\cite{Li:2019xxz} & 51.5 & 50.2\\
            \hline
            $L(n_c)$ & $43.7\pm7.8$~\cite{Zhang:2014yfa} & 53.2 & 54.2 \\
            \hline
            $L(n_0)$ & $52.5\pm17.5$~\cite{Lattimer:2012xj} & 85.9 & 68.3 \\
            \hline
            $J_0$ & $-700\pm500$~\cite{Farine:1997vuz} & $-767$ & $-599$ \\
            \hline
            \hline
        \end{tabular}
\end{table}

\begin{table}[tbh]\small
    \centering
    \caption{The estimation of parameters for bsHLS-L and bsHLS-H.}
\label{tab:para}
        \begin{tabular}{@{}cccccccc}
            \hline
            \hline
            & \(M_{\sigma}(10^5\rm MeV^2)\) & \(\beta'\) & \(r\) & \(g_{\omega NN}\) & \(g_{\rho NN}\) & \(g^{SSB}_{\omega NN}\) & \(g^{SSB}_{\rho NN}\)\\
            \hline
            bsHLS-L & 1.05 & 0.395 & 0.161 & 11.5 & 3.78 & 16.3 & 9.45\\
            \hline
            bsHLS-H & 2.30 & 1.15 & 0.191 & 11.0 & 4.17 & 8.85 & 4.85\\
            \hline
            \hline
        \end{tabular}
\end{table}

To illustrate the features of bsHLS, we consider a Walecka-type model for comparison
\be
\label{eq:linearL}
\mathcal{L}_{\rm RMF} & = & \bar{\psi}\left[i \gamma_\mu \partial^\mu-m_N-g_\sigma \sigma-g_{\omega_{N N}} \gamma_\mu \omega^\mu-g_{\rho_{N N}} \gamma_\mu \rho^{\mu a} \tau^a-g_{\delta}\delta^{a}\tau^{a}\right] \psi \nonumber\\
& &{} +\frac{1}{2} \left(\partial_\mu \sigma \partial^\mu \sigma-m_\sigma^2 \sigma^2\right) -\frac{1}{3} g_2 \sigma^3-\frac{1}{4} g_3 \sigma^4 \nonumber\\
& &{} -\frac{1}{2 g^2} \operatorname{Tr}\left(V_{\mu \nu} V^{\mu \nu}\right)+\frac{1}{2} m_\omega^2 \omega_\mu \omega^\mu+\frac{1}{4} c_3\left(\omega_\mu \omega^\mu\right)^2+\frac{1}{2} m_\rho^2 \rho^{\mu a} \rho_\mu^a\nonumber\\ 
& &{} +\frac{1}{2}\Lambda_V\rho^a_{\mu}\rho^{\mu a}\omega_{\nu}\omega^{\nu}+\frac{1}{2}\left(\partial_{\mu}\delta^a\partial^{\mu}\delta^a-m^2_{\delta}\delta^a\delta^a\right)+\frac{1}{2}C_{\delta\sigma}\sigma^2\left(\delta^a\right)^2\ ,
\ee
where \(\sigma\), \(\omega\), \(\rho\) and \(\delta\) mesons are introduced.
The choices of parameters from some references are listed in Table~\ref{tab:linearM}.
\begin{table}[tbh]\small
    \centering
    \caption{The choice of parameters for Walekca-type models in Eq.~\eqref{eq:linearL}.}
\label{tab:linearM}
        \begin{tabular}{@{}ccccc}
            \hline
            \hline
            &  L-HS~\cite{Horowitz:1981xw} & NL1~\cite{Reinhard:1986qq} & TM1~\cite{Sugahara:1993wz} & FSU-\(\delta 6.7\)~\cite{Li:2022okx} \\
            \hline
            \(n_0({\rm fm^{-3}})\) & 0.149 & 0.152 & 0.145 & 0.148\\
            \hline
            \(m_N\)(MeV) & 939 & 938 & 938 & 938 \\
            \hline
            \(m_{\sigma}\)(MeV) & 520 & 492 & 511 & 492 \\
            \hline
            \(m_{\omega}\)(MeV) & 783 & 795 & 783 & 783\\
            \hline
            \(m_{\rho}\)(MeV) & 770 & 763 & 770 & 763\\
            \hline
            \(g_{\sigma}\) & 10.5 & 10.1 & 10.0 & 10.2\\
            \hline
            \(g_{\omega NN}\) & 13.8 & 13.3 & 12.6 & 13.4\\
            \hline
            \(g_{\rho NN}\) & 4.04 & 4.98 & 4.63 & 7.27\\
            \hline
            \(g_2(\rm fm^{-1})\) & 0 & -12.2 & -7.23 & -8.09\\
            \hline
            \(g_3\) & 0 & -36.3 & 0.618 & 5.88\\
            \hline
            \(c_3\) & 0 & 0 & 71.3 & 172\\
            \hline
            \(g_{\delta}\) & 0 & 0 & 0 & 6.70\\
            \hline
            \(\Lambda_V\) & 0 & 0 & 0 & 204\\
            \hline
            \(m_{\delta}\)(MeV) & 0 & 0 & 0 & 980\\
            \hline
            \(C_{\delta\sigma}\) & 0 & 0 & 0 & 180\\
            \hline
            \hline
        \end{tabular}
\end{table}

\subsection{Nuclear matter properties}

The properties of NM is analyzed at first.
From Table~\ref{tab:nuclear-matter}, one can see that, with appropriate choices of parameters in Table~\ref{tab:para}, our bsHLS can yield NM properties around (sub)saturation density, satisfying the constraints from empirical data.
\begin{figure}[tbh]
    \centering
    \subfigure[Symmetry energy as a function of density.]{
        \includegraphics[width=0.430\textwidth]{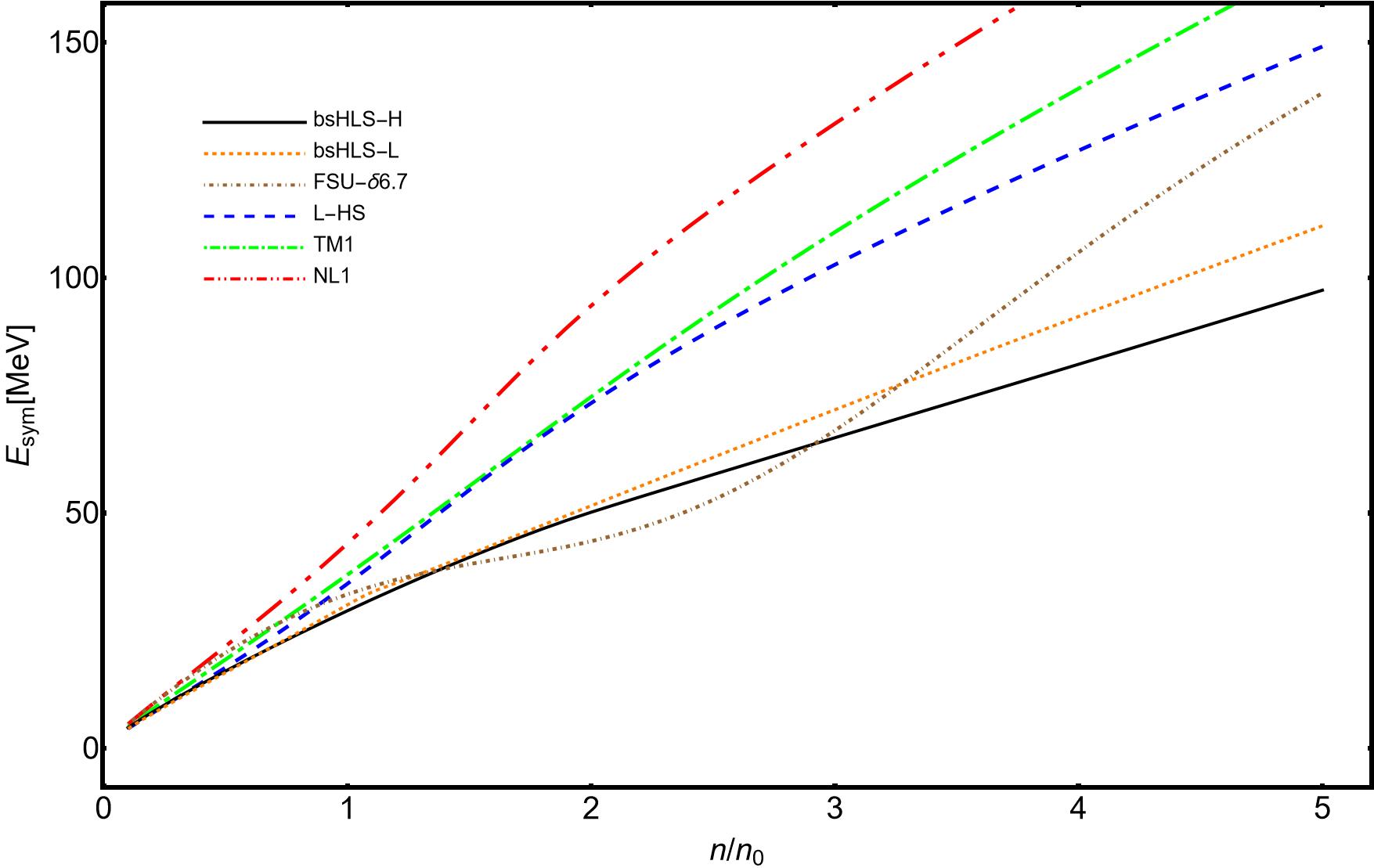}
    }
    \subfigure[Incompressibility as a function of density.]{
        \includegraphics[width=0.455\textwidth]{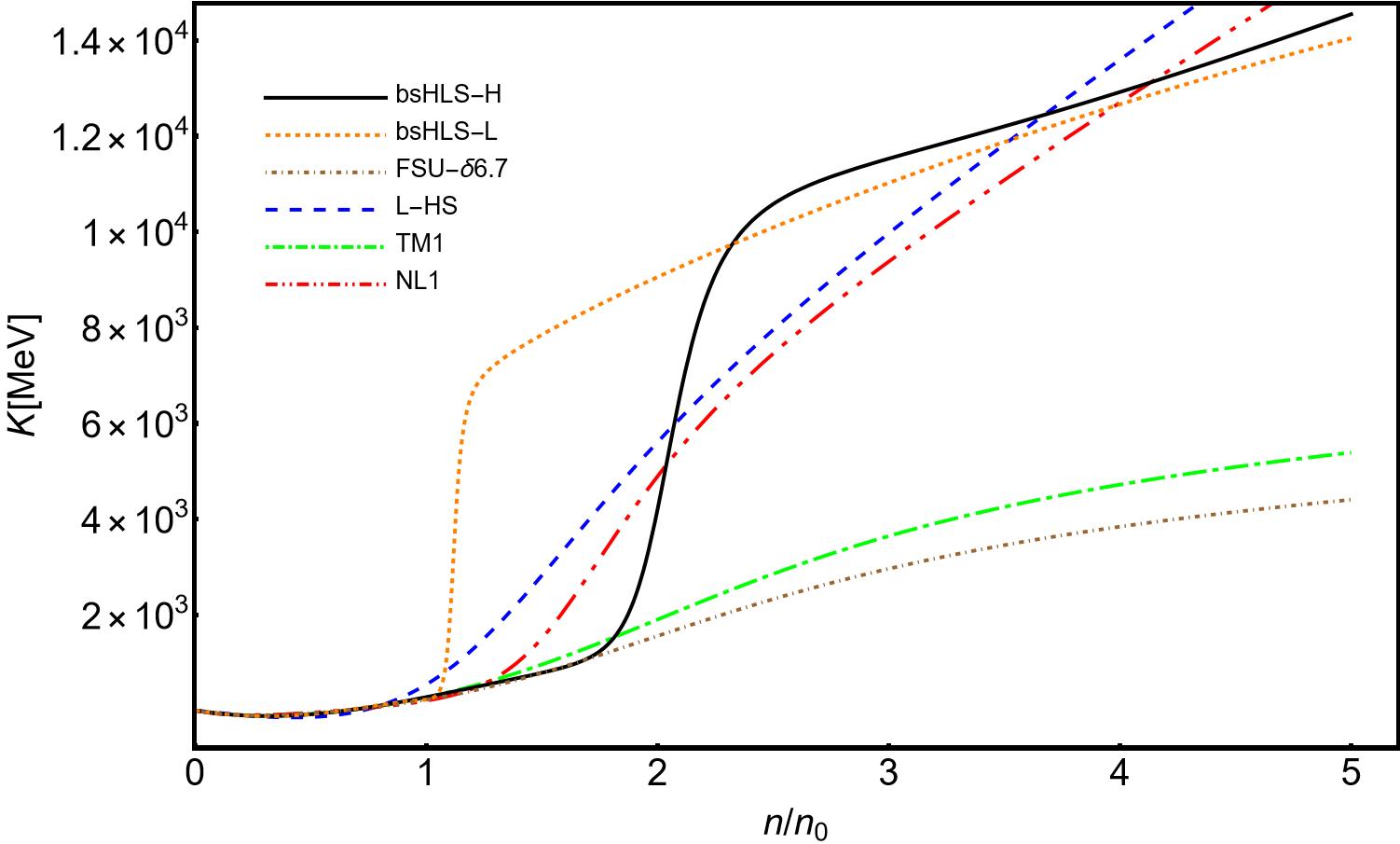}
    }
    \caption{Incompressibility, \(K=9\frac{{\rm d}P}{{\rm d}n}\), which is reduced to \(K_0\) at \(n_0\), and symmetry energy of symmetric NM from bsHLS and Walecka-type models.}~\label{fig:NM}
\end{figure}

Regarding the \(L(n_c)\) results listed in Table~\ref{tab:nuclear-matter}, both bsHLS-L and bsHLS-H exhibit the stiffness to align with the neutron skin thickness of \(\rm Pb^{208}\)~\cite{Reed:2021nqk}, as well as the Walecka-type models compared in this work, shown in Fig.~\ref{fig:NM}.
Moreover, the \(E_{\rm sym}\) behavior of bsHLS is quite similar to that of the compared Walecka-type models at low densities \(n\lesssim n_0\).
However, at intermediate densities, around \(2n_0\), the models can be categorized into two sets: bsHLS and FSU-\(\delta6.7\) yield soft \(E_{\rm sym}\), whereas L-HS, NL1 and TM1 give stiff one.
The Walecka-type models without the \(\delta\) meson are too stiff across the entire density range, while bsHLS provides a more reasonable behavior without introducing \(\delta\).
It will be seen later that this difference can affect the tidal deformation of NS.

For the incompressibility, the results of bsHLS and the compared Walekca-type models show significant differences at intermediate densities:
Walecka-type models exhibit a simple behavior with density, whereas bsHLS shows a kink behavior around \((1\sim2)n_0\).
This kink behavior results in a peak structure in the sound velocity, and is attributed to manifestation of scale symmetry in nuclear matter~\cite{zhang2024peaksoundvelocityscale}.

\subsection{Neutron star structures}

\begin{figure}[tbh]
    \centering
    \includegraphics[width=0.45\textwidth]{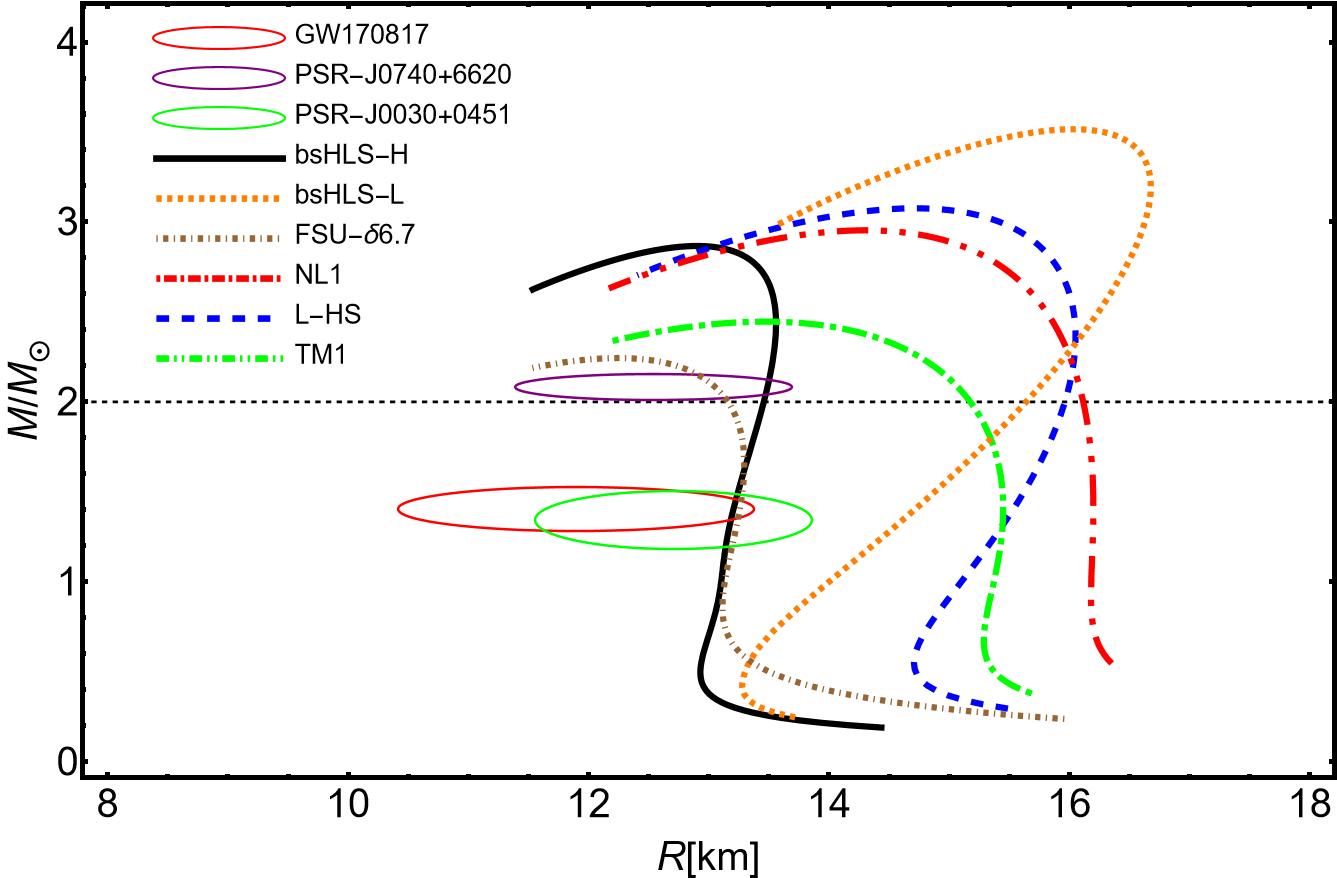}
    \caption{
        The M-R relations from bsHLS and Walecka-type models.
        The constraints are estimated from Refs.~\cite{LIGOScientific:2018cki,LIGOScientific:2018hze,Riley:2019yda,Riley:2021pdl}.
        The M-R relation is calculated by solving the Tolman-Oppenheimer-Volkoff (TOV) equation~\cite{Oppenheimer:1939ne,Tolman:1939jz}.
        All EoSs for TOV equation calculation are interpolated to a BPS EoS~\cite{Baym:1971pw} below \(0.01\ \rm fm^{-3}\)~\cite{Arnett:1977czg}.
    }~\label{fig:MR}
\end{figure}

Next, the NS structures are studied using the EOSs discussed above for pure neutron matter (PNM). 
The results of NS structures are shown in Fig.~\ref{fig:MR} and Table~\ref{tab:td}.

\begin{table}[tbh]\small
    \centering
    \caption{
        Tidal deformations from bsHLS and Walecka-type models defined in Eq.~\eqref{eq:linearL} and Table~\ref{tab:linearM} for NS with mass of \(1.4\ M_{\odot}\).
        They are calculated from the EOSs used for the M-R relation in Fig.~\ref{fig:MR}, using the formalisms in Ref.~\cite{Postnikov:2010yn}.
    }
\label{tab:td}
        \begin{tabular}{@{}ccccccc}
            \hline
            \hline
            & bsHLS-L & bsHLS-H & TM1 & L-HS & NL1 & FSU-\(\delta6.7\) \\
            \hline
            \(\Lambda_{1.4}\) & 2120 & 910 & 2240 & 2780 & 2620 & 878 \\
            \hline
            \hline
        \end{tabular}
\end{table}
From Fig.~\ref{fig:MR}, one can see that the mass-radius (M-R) relation obtained from bsHLS-H falls within the constraints of astro-observations.
Although the result from FSU-\(\delta6.7\) satisfies these constraints, it requires introducing a new degree of freedom \(\delta\).
The Walecka-type models without \(\delta\) considered in this work cannot provide results within the constraints of NSs, though they reproduce the NM properties around \(n_0\)~\cite{Horowitz:1981xw,Reinhard:1986qq,Sugahara:1993wz}.
This highlights the advantage of bsHLS in interpreting the structures of nuclei and NSs within a unified framework.
It should be emphasized that the present analysis indicates that, in order to have realistic parameter space of nuclear model, including Walecka-type models, the data of both NM properties and NS structures should be considered in the statistical analysis simultaneously, as discussed in Ref.~\cite{Guo:2023mhf}.
When meeting the NS constraints, the \(M_{\rm max}\) of NSs predicted by these Walecka-type models and chiral nuclear force models is around or slightly above \(2M_{\odot}\), as discussed in Refs.~\cite{Ozel:2016oaf, Guo:2023mhf,Fan:2023spm}.
This makes bsHLS more appealing, as the predicted \(M_{\rm max}\) of bsHLS-H is \(\sim 2.8M_{\odot}\).
This predicted \(M_{\rm max}\) may have profound implications for addressing the gap problem in the continuous mass distribution of supernova remnants~\cite{Bailyn:1997xt, Farr:2010tu} and for understanding the gravitational wave event GW190814~\cite{LIGOScientific:2020zkf}.

For tidal deformations from Table~\ref{tab:td}, one can conclude that only the results of bsHLS-H and FSU-\(\delta6.7\) are close to the constraints of GW170817~\cite{LIGOScientific:2018cki}.
This is due to the softness of \(E_{\rm sym}\) from these two models at intermediate densities, as shown in Fig.~\ref{fig:NM}.
Since \(E_{\rm sym}\) from bsHLS-L is stiffer than bsHLS-H and FSU-\(\delta6.7\) below \(\simeq 3n_0\), bsHLS-L gives larger \(\Lambda_{1.4}\).
The same reasoning applies to the Walecka-type models, NL1, L-HS, and TM1.

In summary, the above discussion indicates that bsHLS-H is a reasonable model for NM and NS.
It implies that the scaling parameter \(r\approx0.19\), \(\beta'\approx1.15\), \(M_{\sigma}=m_{\sigma}f_{\chi}\approx 2.3\times10^{5}~\rm MeV^2\), and the couplings between vector mesons and nucleons \(g_{\rho NN}\approx4.17\), \(g_{\omega NN}\approx 11.0\).
The scaling parameter \(r\approx0.19\) is consistent with pion-nuclei bound state data~\cite{Kienle:2004hq}.
And the \(\beta'\approx1.15\) agrees with estimates from skyrmion crystal approach~\cite{Ma:2016nki,Shao:2022njr}.
If \(f_{\chi}\) is taken to be \(3f_{\pi}\approx270~\rm MeV\), \(m_{\sigma}\approx 850~\rm MeV\), while the power-counting mechanism of Crewther and Tuntall remains valid~\cite{Crewther:2013vea}.
The coupling constant \(g_{\rho NN}\approx4.17\) aligns with the results from the OBE potential analysis of nucleon interactions~\cite{Erkelenz:1974uj}, and \(g_{\omega NN}\approx 11.0\) is in agreement with the analyses of nucleon-nucleon scatterings~\cite{Backman:1984sx}.

\subsection{The patterns of scale symmetry and phenomenologies}

The difference between bsHLS-L and bsHLS-H becomes significant in the M-R relation, as shown in Fig.~\ref{fig:MR}, though it is not distinguishable from the NM properties around saturation density.
The key difference lies in \(\beta'\), which is connected to the existence of the pseudo-conformal structure at high densities~\cite{Shao:2022njr}. 
The \(\beta'\) of bsHLS-H aligns with the constraints of the pseudo-conformal limit, whereas the \(\beta'\) of bsHLS-L does not.
This highlights the impact of scale symmetry patterns on NS structures.

\begin{figure}[tbh]
    \centering
    \includegraphics[width=0.45\textwidth]{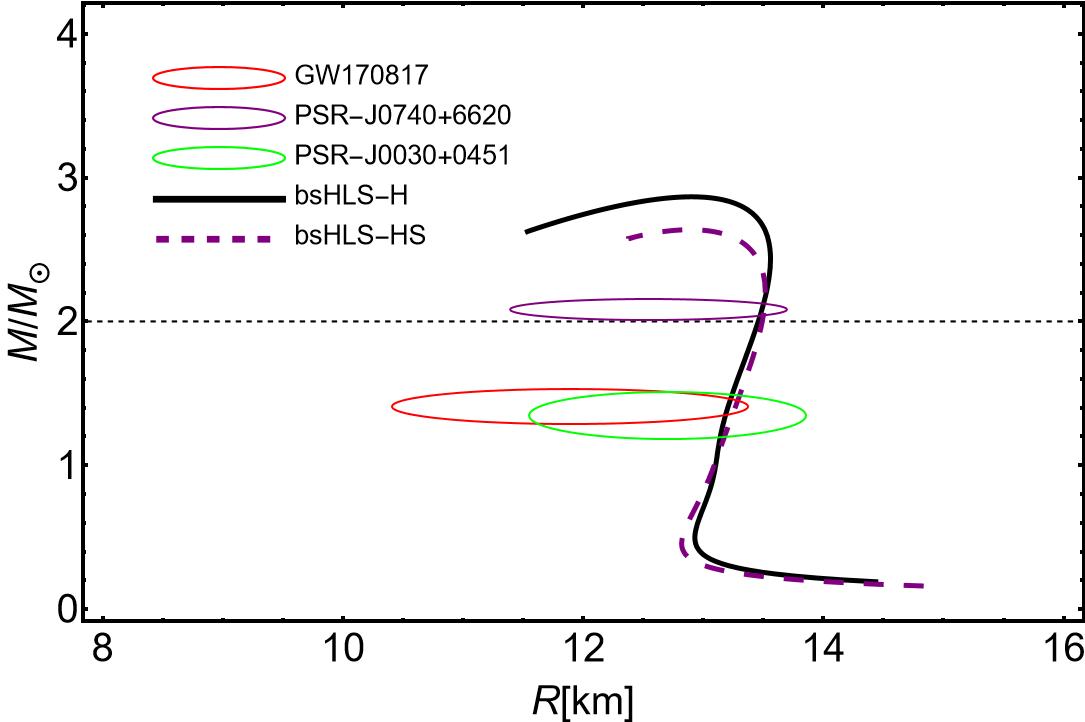}
    \caption{
        NS structure results with/without $g_{\omega NN}$ suppression.
    }~\label{fig:MRsup}
\end{figure}

Moreover, since $\sigma$ is nonlinearly coupled with other mesons (see Eq.~\eqref{eq:EOMv}) through a conformal compensator, its density dependence shows a kink behavior that is not observed in Walecka-type models~\cite{zhang2024peaksoundvelocityscale}. 
As the result, the order parameter $\langle\chi\rangle^*$, calculated based on bsHLS does not approach zero throughout the density regions, which is necessary for the (pseudo-)conformal limits of QCD at high densities~\cite{Borsanyi:2013bia, Ma:2018qkg, Shao:2022njr}. 
To recover the expected behavior of $\langle\chi\rangle^*$, a possible approach is to couple an additional factor to $g_{\omega NN}$ to obtain the effective coupling $\tilde{g}_{\omega NN}=g_{\omega NN}/(1+R\frac{\rho_p+\rho_n}{n_0})$~\cite{Paeng:2013xya,zhang2024peaksoundvelocityscale}.
It is found that, as listed in Table~\ref{tab:nm-sup}, the NM properties can be reproduced with the parameter set bsHLS-HS:
\(M_{\sigma}=2.25\times10^5\ \rm MeV^2\), \(\beta'=1.14\), \(g_{\omega NN}=11.5\), \(g_{\rho NN}=4.27\), \(g^{SSB}_{\omega NN}=8.70\), \(g^{SSB}_{\rho NN}=4.85\), \(r=0.20 \) and \(R=0.02 \).
\begin{table}[tbh]\small
    \centering
    \caption{
        The quantities of nuclear matter with additional suppressions of \(\tilde{g}_{\omega NN}\), and the definitions and constraints are the same as Table~\ref{tab:nuclear-matter}.
    }
    \begin{threeparttable}~\label{tab:nm-sup}
        \begin{tabular}{@{}cccccccccc}
            \hline
            \hline
            & $n_0$ & \(e_0\) & $K_0$ & $E_{\mathrm{sym}}(n_c)$ & $E_{\mathrm{sym}}(n_0)$ & $E_{\mathrm{sym}}(2n_0)$ & $L(n_c)$ & $L(n_0)$ & $J_0$ \\
            \hline
            bsHLS-HS & 0.159 & -16.0 & 259 & 21.6 & 30.3 & 52.9 & 55.4 & 74.5 & -720 \\
            \hline
            \hline
        \end{tabular}
    \end{threeparttable}
\end{table}
And, as shown in Fig.~\ref{fig:MRsup}, the M-R relations for PNM are roughly consistent with observational constraints, but there is a decrease of $M_{\rm max}$ compared to bsHLS-H.

\begin{figure}[tbh]
    \centering
    \subfigure[Imcompressibilities as a function of density in symmetric NM.]{
        \includegraphics[width=0.45\textwidth]{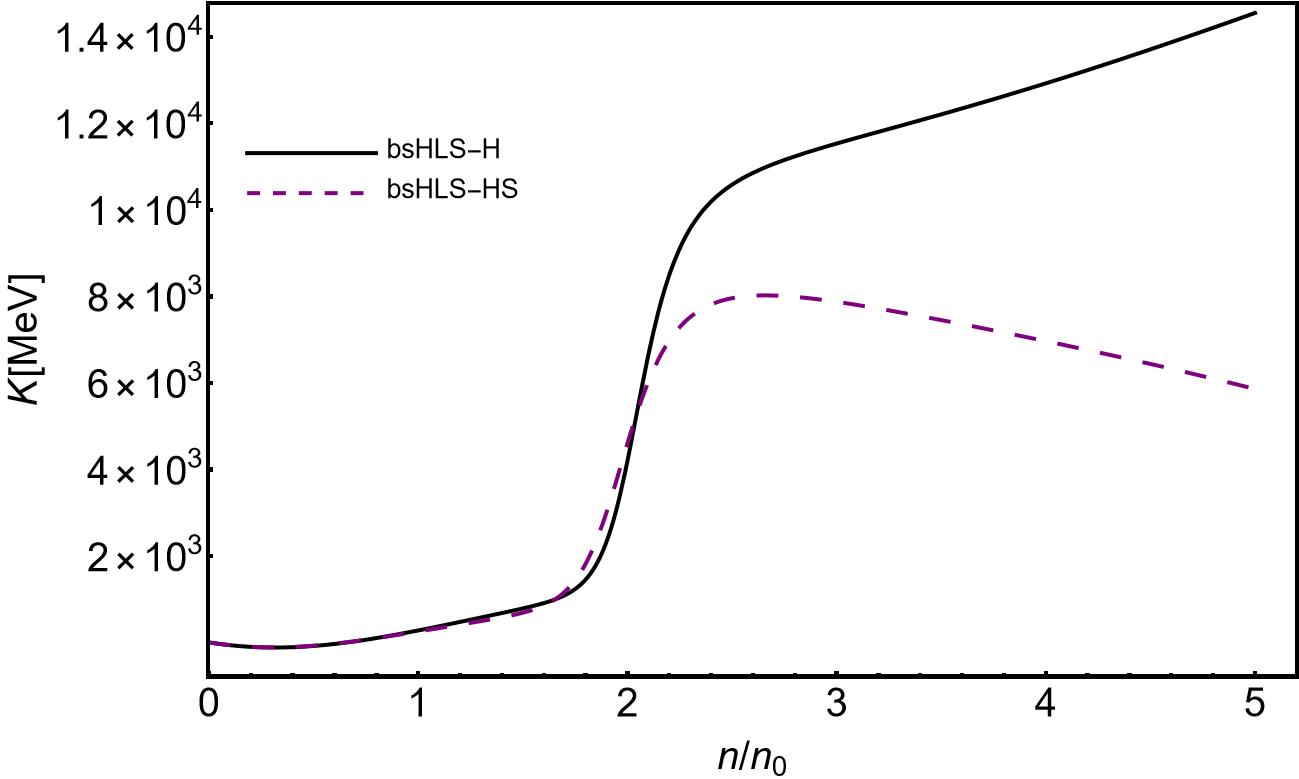}
    }
        \subfigure[Symmetry energies as a function of density.]{
        \includegraphics[width=0.43\textwidth]{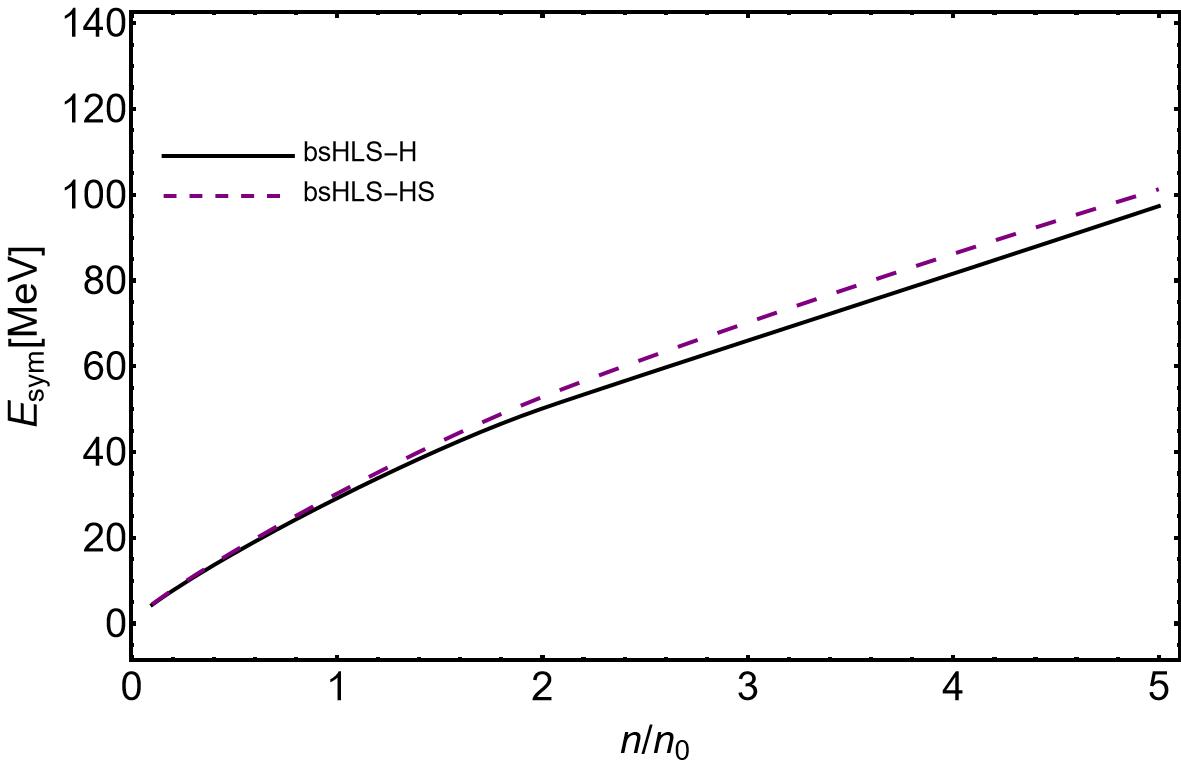}
    }
    \caption{
        NM property results with/without $G_\omega$ suppression.
    }~\label{fig:NMsup}
\end{figure}

In order to understand the discrepancies in the M-R relations, the incompressibilities and symmetry energies are also calculated for bsHLS-H and bsHLS-HS, shown in Fig.~\ref{fig:NMsup}.
It can be seen that the symmetry energy remains almost the same in both cases.
However, the incompressibility of bsHLS-HS is much softer than that of bsHLS-H at high densities, leading to a decrease in \(M_{\rm max}\).
At low or intermediate densities, the incompressibilities of both cases are similar, which results in the M-R relations being comparable in the region of NS constraints.

\begin{figure}[tbh]
    \centering
    \subfigure[$\langle\chi\rangle^*/f_{\chi}$ as a function of density.]{
        \includegraphics[width=0.45\textwidth]{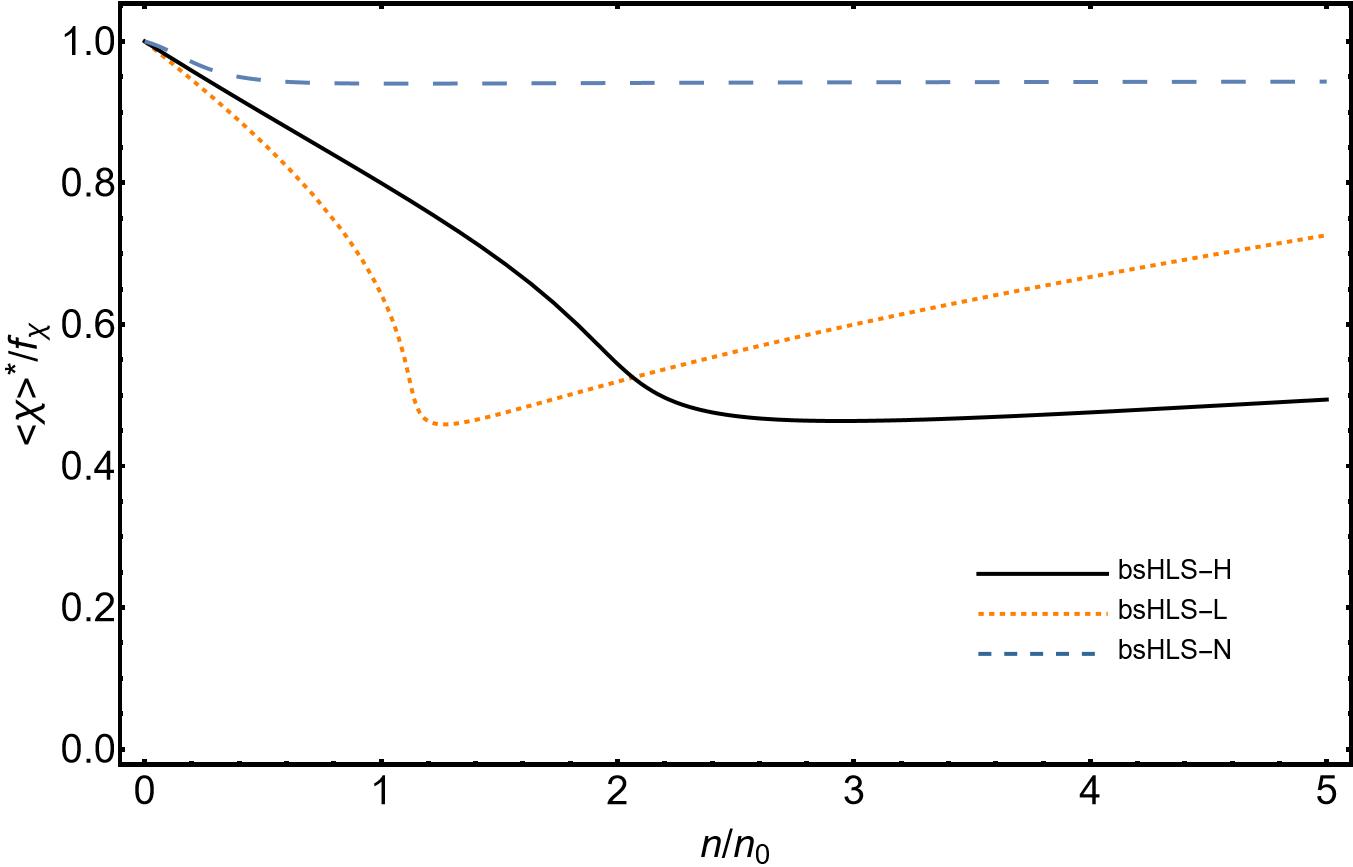}
    }
        \subfigure[MR relation of NS.]{
        \includegraphics[width=0.44\textwidth]{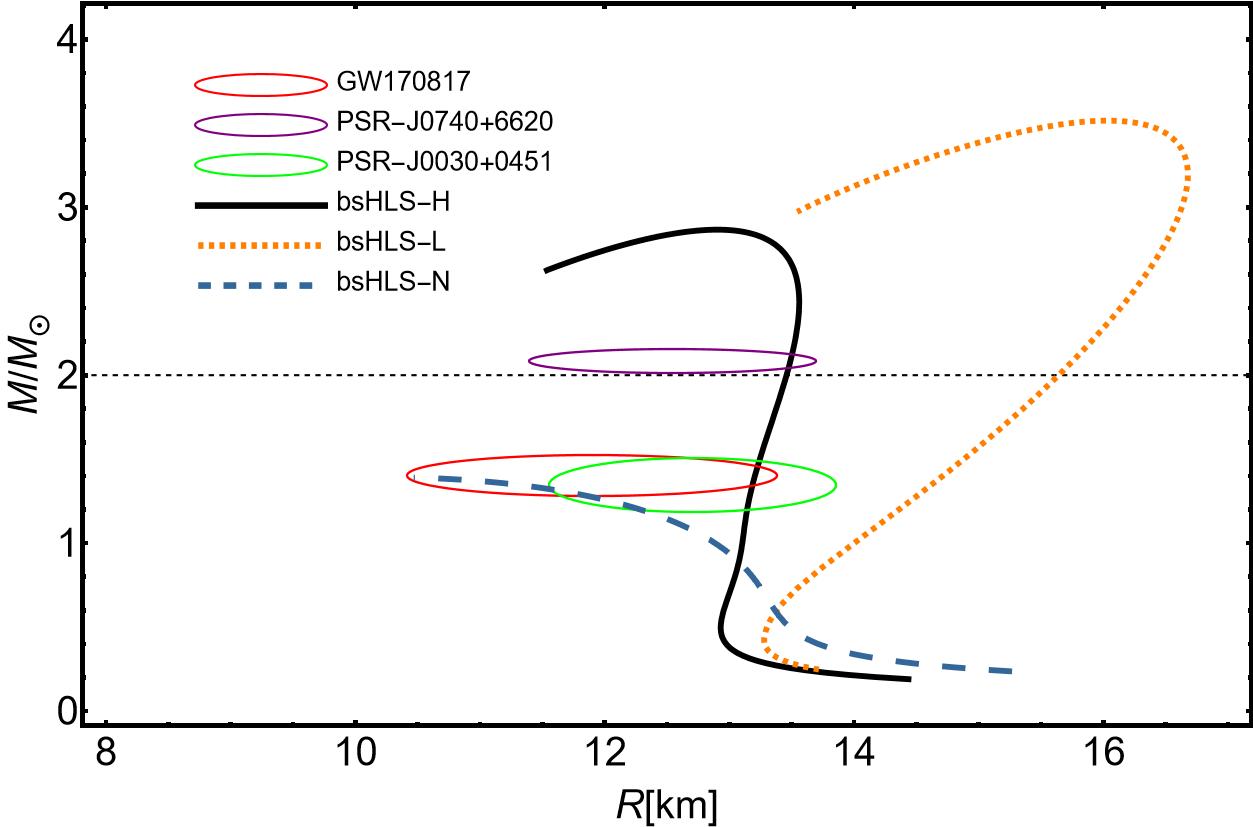}
    }
    \caption{
        The $\langle\chi\rangle^*/f_{\chi}$ in pure neutron matter and NS structure results with/without B-R scaling.
    }~\label{fig:NBR}
\end{figure}

Additionally, the NM properties are investigated with B-R scaling turned off.
We find that with the parameter set bsHLS-N: \(M_{\sigma}=1.11\times10^5\ \rm MeV^2\), \(\beta'=1.10\), \(g_{\omega NN}=-11.0\), \(g_{\rho NN}=4.17\), \(g^{SSB}_{\omega NN}=-192\) and \(g^{SSB}_{\rho NN}=4.85\), the NM properties can still be reproduced (e.g., $n_0=0.16\ \rm{fm}^{-3}$, $e_0=-16.0\ \rm{MeV}$, $K_0=241\ \rm{MeV}$, $E_{{\rm sym}}(n_0)=29.3\ \rm{MeV}$).
However, this approach results in a much softer scale symmetry parameter $\langle\chi\rangle^*$ flow compared to the cases with B-R scaling.
As a result, the M-R relation becomes unnatural and deviates from the constraints, as shown in Fig.~\ref{fig:NBR}.

All these findings, including the necessity of introducing $\tilde{g}_{\omega NN}$ suppression and B-R scaling, signify the importance of properly parametrizing the scale symmetry in densities.
This reminds us of the past work on the quenching factors of $g_A$ which affect the $\beta$ decay of neutrons in dense environments (see Ref.~\cite{Ma:2020tsj} for details).
From the comparison among the various cases discussed above, valuable lessons about NS and NM properties have been learned, providing guidance on hadron interaction parameterization to describe dense NM.
Since the differences between bsHLS and Walecka-type models in our analysis are at intermediate density regions, not far from \(n_0\), it is expected that examining bsHLS in dense systems will be promising in future experiments.

\section{Conclusion and discussion}

\label{sec:Conc}

In this work, the bsHLS, constructed based on the philosophy of \(\chi \rm EFT\) with HLS and a possible IR fixed point of QCD at low energies, is applied to dense environments using the RMF approximation, with free parameters fixed by pinning the nuclei structure data around \(n_0\).
The NM properties and NS structures can be well reproduced, closely matching empirical values in bsHLS-H case.

Without introducing many freedoms, such as the \(\delta\) meson, and with operators organized to respect chiral and scale symmetry considerations and expanded by chiral-scale orders, the bsHLS can provide a reasonable behavior of NM properties from subsaturation to intermediate densities, e.g., \(K(n)\) and \(E_{\rm sym}(n)\), compared to Walecka-type models.
And the NS structures are sensitive to NM properties at these density regions, making bsHLS outperform Walecka-type models in describing a wider range of densities.
More specifically, the kink behavior of the \(\sigma\) field in bsHLS at intermediate densities allows the \(M_{\rm max}\) to reach nearly \(3\ M_{\odot}\) for PNM, while other NS observational constraints are still statisfied.

Besides, the behaviors of symmetry patterns in dense environments are also found to be pivotal to macroscopic phenomena:  If there is no restoration point of scale symmetry at certain densities, such as the \(\beta'\) value of bsHLS-L, the NS structures will fall outside observational constraints; The \(M_{\rm max}\) of predicted NSs is influenced by the behavior of order parameter of scale symmetry, \(\langle\chi\rangle^*\).
Furthermore, the study on the flow of \(\langle\chi\rangle^*\) with densities suggests the necessity of introducing an additional suppression factor for $\tilde{g}_{\omega NN}$, and it could be an interesting problem for further investigation.

In summary, introducing bsHLS to NM studies is a promising approach due to its close relation to QCD symmetry patterns and the effective potentials organized by chiral-scale orders, which have already proven successful in describing scattering experiments at vacuum.
Furthermore, the difference between bsHLS and Walecka-type models is not far from \(n_0\), making it possible to be verified in future experiments, such as heavy ion collisions.
The relationship between microscopic symmetries and macroscopic phenomena found in this work is also a valuable topic to be further studied.

\section*{Acknowledgments}

We would like to thank Yi-Zhong Fan for his valuable comments.

The work of Y.~L. M. is supported in part by the National Science Foundation of China (NSFC) under Grant  No. 12347103 and No. 11875147.



\bibliography{RefChiralScaleMF}

\end{document}